\documentclass[preprint,preprintnumbers,amsmath,amssymb]{revtex4}

\usepackage{graphicx}% Include figure files
\usepackage{bm}% bold math

\begin{document}

\preprint{USTC-ICTS-08-22}

\title{Holographic Gas as Dark Energy}

\author{Miao Li$^{1,2}$, Xiao-Dong Li$^{1}$, Chunshan Lin$^{1}$, Yi Wang$^{2,1}$}

\affiliation{%
$^1$Interdisciplinary Center for Theoretical Study, USTC, Hefei,
Anhui 230026, P.R.China\\$^2$Institute of Theoretical Physics, CAS,
Beijing 100080, P.R.China
\\\vspace{1cm}
}%

%\date{\today}% It is always \today, today,
             %  but any date may be explicitly specified

\begin{abstract}
We investigate the statistical nature of holographic gas, which may
represent the quasi-particle excitations of a strongly correlated
gravitational system. We find that the holographic entropy can be
obtained by modifying degeneracy. We calculate thermodynamical
quantities and investigate stability of the holographic gas. When
applying to cosmology, we find that the holographic gas behaves as
holographic dark energy, and the parameter $c$ in holographic dark
energy can be calculated from our model. Our model of holographic
gas generally predicts $c<1$, implying that the fate of our
universe is phantom like.
\end{abstract}

\maketitle

\section{Introduction}
The dark energy problem \cite{ccref} has become one of the most
outstanding problems in cosmology. Dark energy should be considered
as a problem of quantum gravity, because it is a problem of both quantum
vacuum fluctuations and the largest scale gravity
of our universe. One opinion believes that our
quasi-de Sitter universe only has finite degrees of freedom, which
is related to the cosmological constant. As the number of degrees of
freedom must take integer value, one concludes that the cosmological
constant should also be quantized. The classical general relativity
does not quantize the cosmological constant, so one must appeal to
quantum gravity \cite{Witten:2001kn}. Moreover, the Newton's
constant appears at the denominator of the entropy formula
$S=\frac{A}{4G}$, indicating that the nature of dark energy is
nonperturbative in the Newton's constant.

Achievements in condensed matter physics show that sometimes a system
which appears nonperturbative can be described by weakly interacting
quasi-particle excitations. We hope that similar mechanisms may work
for gravity. In this paper, we investigate phenomenologically a gas
of holographic particles, which we hope represents the
quasi-particle excitations of a strongly correlated gravitational
system. The statistical properties of this holographic gas is
studied in Sec. \ref{thermo}. In Sec. \ref{Stability}, we consider
the stability of holographic gas.

We apply holographic gas to cosmology in Sec. \ref{cosmology}. We
find interestingly that energy density of holographic dark energy
\cite{Li:2004rb}\cite{refHDE} emerges once we assume that the volume
of holographic gas is bounded by the future event horizon. Our
investigation of the holographic gas is independent of the vacuum
energy bound
 proposed by Cohen, Kaplan and Nelson \cite{Cohen:1998zx}, which is the motivation of holographic dark
energy. So the connection seems nontrivial, and may represent some
underlying features of quantum gravity. We also find that the
parameter $c$, which is left as a free parameter, can be calculated
within the holographic gas framework. Holographic gas generally
predicts $c^2<1$, which represents a phantom like dark energy at
late times. The parameter region $c^2 \geq 1$ can also be produced
with some {\it ad hoc} assumptions.

\section{Thermodynamics of Holographic Gas}\label{thermo}

It is believed that the entropy of our quasi-de Sitter universe is
proportional to the area of some cosmic horizon. If this
quasi-de Sitter entropy is realized by particles in the bulk, then
the entropy of this ``holographic gas'' should be proportional to
the boundary area. In this section, we seek statistical systems of
particles with this property.

We assume relativistic dispersion relation $\epsilon=k$ for a
holographic particle. We will also give the result for modified
dispersion relation in this section, leaving the derivation to
Appendix A. We consider simultaneously indistinguishable
Maxwell-Boltzmann statistics, Bose-Einstein statistics and
Fermi-Dirac statistics. As will be shown, the results of these three
statistics only differ by a constant, which drops out when applying
to cosmology.

We assume that the degeneracy of a holographic particle with fixed momentum depends on the momentum of the particle and/or the spatial volume,
\begin{equation}
  \omega =\omega_0 k^aV^bM_p^{3b-a}~,
\end{equation}
where $\omega_0$ is a dimensionless constant, and $M_p=(8\pi
G)^{-1/2}$ is the reduced Planck mass.

We use ${\cal Z}$ to denote the grand partition function of
holographic gas, which is a function of temperature $T$, volume $V$
and chemical potential $\mu$. In the Maxwell-Boltzmann case, the single
particle canonical partition function is
\begin{equation}
  Z_1 = \sum_{i} e^{-\epsilon_i/T}= V\int \frac{d^3k}{(2\pi)^3} ~\omega ~ e^{-k/T}= \frac{\omega_0}{2\pi^2}\Gamma(a+3)M_p^{3b-a}V^{1+b}T^{a+3}~,
\end{equation}
where the summation variable $i$ runs over all possible states, and $\Gamma(x)$ is the Gamma function. If interaction between holographic particles can be neglected, the grand partition can be written as
\begin{equation}
  {\cal Z}_{\rm MB}=\sum_{N=0}^{\infty}\frac{1}{N!}e^{\mu N/T}Z_1^N=\exp \left\{e^{\mu/T}Z_1\right\}~,
\end{equation}
where $N!$ appears because the particles are indistinguishable.

The logarithm of the grand partition function can be calculated as
\begin{equation}\label{ZMB}
  \ln {\cal Z}_{\rm MB}= e^{\mu/T}Z_1 = \frac{\omega_0}{2\pi^2}\Gamma(a+3)M_p^{3b-a} e^{\mu/T}V^{1+b}T^{a+3}~.
\end{equation}

For the Bose-Einstein and Fermi-Dirac cases, we have
\begin{align}
  {\cal Z}_{\rm BE}=\prod_{i=1}^{\infty} \frac{1}{1-\exp\{(\mu-\epsilon_i)/T\}}~,~~~  {\cal Z}_{\rm FD}=\prod_{i=1}^{\infty} \left(1+\exp\{(\mu-\epsilon_i)/T\}\right)~
\end{align}
The logarithm of the partition function can be integrated out as
\begin{align}\label{ZBEFD}
  \ln {\cal Z}_{\rm BE}&= -V\int \frac{d^3k}{(2\pi)^3}\omega \ln \left(1-\exp\left\{\frac{\mu-k}{T}\right\}\right) = {\rm Li}_{a+4}(e^{\mu/T}) Z_1~,\nonumber\\
  \ln {\cal Z}_{\rm FD}&= V\int \frac{d^3k}{(2\pi)^3}\omega \ln \left(1+\exp\left\{\frac{\mu-k}{T}\right\}\right)= -{\rm Li}_{a+4}(-e^{\mu/T})Z_1~,
\end{align}
where $\rm Li_{n}(z)\equiv \sum_{N=1}^{\infty}z^N/N^n$ is the polylogarithm function. In order that $\ln{\cal Z}$ is finite in Eqs. \eqref{ZMB} and \eqref{ZBEFD}, we require $a>-3$. For cosmological usage in Sec. \ref{cosmology}, we note that when $\mu=0$, the polylogarithm function reduces to the Riemann zeta function,
\begin{align}\label{ZBEFDm0}
  \ln {\cal Z}_{\rm BE}&= \zeta(a+4) \ln {\cal Z}_{\rm MB}~,\nonumber\\
  \ln {\cal Z}_{\rm FD}&= \left(1-2^{-a-3}\right)\zeta(a+4)\ln {\cal Z}_{\rm MB}~.
\end{align}

For all the three kinds of statistics, the total energy and total entropy of holographic gas within volume $V$ take the form
\begin{equation}\label{ES}
  E=T^2\partial_T \ln {\cal Z}+\mu T \partial_\mu \ln {\cal Z}=(a+3)T\ln {\cal Z}~,~~~S=E/T+\ln {\cal Z}=(a+4)\ln {\cal Z}~.
\end{equation}
and the particle number $N$ is
\begin{equation}
  N_{\rm MB}=\ln {\cal Z}_{\rm MB}~,\qquad N_{\rm BE}=\frac{{\rm Li}_{a+3}(e^{\mu/T})}{{\rm Li}_{a+4}(e^{\mu/T})}\ln {\cal Z}_{\rm BE}~,\qquad N_{\rm FD}=\frac{{\rm Li}_{a+3}(-e^{\mu/T})}{{\rm Li}_{a+4}(-e^{\mu/T})}\ln {\cal Z}_{\rm FD}~.
\end{equation}

Inspired by holography, we take $T\propto V^{-1/3}$, and require
that $S$ is proportional to the area of the system. In terms of Eqs.
\eqref{ZMB}, \eqref{ZBEFD} and \eqref{ES}, this requirement is
reduced to a relation between $a$ and $b$,
\begin{equation}
  3b-a=2~.
\end{equation}
The energy density $\rho$ can be written as
\begin{equation}\label{rho1}
  \rho=\frac{E}{V}=\frac{a+3}{a+4}\frac{ST}{V}~.
\end{equation}

If we allow modified dispersion relation $\epsilon= \epsilon_0 M_p^{3n-m}p^mV^n$, we find in Appendix A that energy density takes the form
\begin{equation}
  \rho=\frac{a+3}{a+3+m}\frac{ST}{V}~,
\end{equation}
and $(a+3)/m >0$ is needed to make the partition function finite.

There is one more possibility that the holographic gas obeys an
exotic kind of statistics named ``infinite statistics''
\cite{infistat}\cite{ng}. The properties of holographic gas obeying
infinite statistics are given in the Appendix C.

At the end of this section, we would like to pause and comment on
the universality of the first law of thermodynamics. Within the
volume $V$ that we consider, the first law of thermodynamics
$dE=TdS-pdV+\mu dN$ is automatically satisfied. This serves as a
consistency check of our calculation. However, the first law of
thermodynamics is not necessarily satisfied if we concentrate on a
subsystem $\tilde V$ inside $V$, with some process $d(\tilde
V/V)\neq 0$. This is because for the holographic gas, physical
quantities such as entropy is not extensive. So Euler's relation
$E=TS-pV+\mu N$ is not satisfied. The violation of Euler's relation
is equivalent to the statement that the first law of thermodynamics
is not satisfied within an arbitrary volume, this is the case of the
holographic gas.

\section{Stability of Holographic Gas}\label{Stability}
In this section, we investigate stability of holographic gas in
three aspects. We show that the physical pressure and the specific
heat of holographic gas are both positive. We also calculate the
mean energy fluctuation, and show that it is very small.

The physical pressure can be calculated as
\begin{equation}
p=T\frac{\partial{\ln {\cal Z}}}{\partial V}=\frac{1+b}{a+3}\rho>0~.
\end{equation}
A positive pressure indicates that holographic gas can not collapse globally.

The specific heat takes the form
\begin{equation}
  C_V=(a+3)(a+4)\ln {\cal Z}>0~.
\end{equation}
For hot holographic gas, it has a high temperature and tends to lose
heat. Because its $C_V$ is positive, it will become colder after
losing heat and ultimately stops losing heat. The situation is
similar for cold holographic gas.

The mean fluctuation $\delta E$ is defined by
\begin{equation}
\delta E^2=\langle E^2 \rangle -\langle E\rangle ^2~,
\end{equation}
where the expect value of energy is
\begin{equation}
\langle E \rangle=\frac{T^2}{{\cal Z}}\frac{\partial}{\partial T}{\cal Z}~,
\end{equation}
and
\begin{equation}
\langle E^2\rangle=\frac{1}{{\cal Z}}\frac{\partial^2}{\partial (1/T)^2}{\cal
Z}=\frac{T^2}{{\cal Z}}\frac{\partial}{\partial T}({\cal
Z}\langle E\rangle)=\langle E\rangle^2-T^2\frac{\partial \langle E\rangle}{\partial T}~.
\end{equation}
So we get
\begin{equation}
\delta E^2=T^2\frac{\partial\langle E\rangle}{\partial T}=(a+3)(a+4) T^2 \ln{\cal
Z}~,
\end{equation}
and
\begin{equation}
\frac{\delta E^2}{E^2}=\frac{a+4}{a+3}\frac{1}{\ln {\cal Z}}~,\qquad \ln{\cal Z}\sim N\gg 1.
\end{equation}
We find the above quantity is very small, which also implies the
stability of the statistical fluctuation of holographic gas model.

Moreover, the typical wave length of the holographic gas is $\lambda
\sim V^{1/3}$. A holographic particle is wave-like within the
volume, and it has nowhere to collapse. We conclude that holographic
gas is stable for the above reasons.

\section{Applications to Dark Energy}\label{cosmology}

\subsection{Physical Pressure and Effective Pressure}
To apply our statistical results to cosmology, the first thing one should take care of is the distinction between the physical pressure and the effective pressure.

The physical pressure is defined as
\begin{equation}
  p=T\frac{\partial{\ln {\cal Z}}}{\partial V}~,
\end{equation}
which satisfies automatically the first law of thermodynamics within the volume $V=4\pi R^3/3$,
\begin{equation}\label{firstlaw}
d\rho + 3\frac{d R}{R}(\rho+p) =\frac{TdS}{4\pi R^3/3}~.
\end{equation}
The physical pressure can be in principle measured in a local experiment such as using a barometer (on condition that one could build a barometer that can interact with the holographic gas).

The effective pressure is defined as
\begin{equation}\label{eff}
  \dot\rho+3H(\rho+p_{\rm eff})=0~,
\end{equation}
where $H=\dot a/a$ is the Hubble parameter. The effective pressure is responsible for the evolution of the universe. It is because the Friedmann equation $3M_p^2 H^2=\rho$ states that the evolution of the universe is governed by the energy density, and it is the effective pressure that controls how energy density evolves when the universe expands.

Dividing Eq. \eqref{firstlaw} by $dt$, and using Eq. \eqref{eff} to cancel $\dot \rho$, we get the relation between the effective pressure and the physical pressure,
\begin{equation}
  p_{\rm eff}=-\rho+\frac{\dot R}{HR}(\rho+p)-\frac{T\dot S}{4\pi HR^3}~.
\end{equation}
However, when applying to cosmology, usually we do not need to really calculate this effective pressure. Because we can use directly energy density to solve the evolution of the universe.

\subsection{Energy Density of Holographic Gas}
 Now let us apply Eq. \eqref{rho1} into cosmology. We use the Gibbons-Hawking entropy $S=8\pi^2R^2M_p^2$ and temperature $T=1/(2\pi R)$, where $R$ is the radius of our universe, which may be taken as the radius of the future event horizon, particle horizon, or apparent horizon. We further assume the chemical potential $\mu=0$, because when applying to cosmology, the number of holographic particles can change when the area of cosmic horizons changes. Eq. \eqref{rho1} takes the form
\begin{equation}\label{rho2}
  \rho=3\frac{a+3}{a+4} M_p^2R^{-2}~.
\end{equation}
It is well known that the particle horizon or the apparent horizon with an energy density of this type can not accelerate the universe. On the other hand, if we take $R$ as the radius of the future event horizon $R_h=a(t)\int_t^{\infty}dt'/a(t')$, compared with the energy density of holographic dark energy $\rho=3c^2M_p^2R_h^{-2}$, we have
\begin{equation}
  c^2=\frac{a+3}{a+4}~,
\end{equation}
In other words, holographic gas reproduces the energy density of
holographic dark energy, and gives statistical interpretation for
the parameter $c$. We note that when $a>-3$, which is required to
have a finite partition function, we always have $c^2<1$. This
hopefully explains why people always get $c^2<1$ in the data fitting
\cite{datafit}. This result also holds with a non-vanishing chemical
potential and modified dispersion relation,
\begin{equation}
  c^2=\frac{a+3}{a+3+m}~,
\end{equation}
So holographic gas predicts rather robustly that dark energy behaves like phantom at late times.

Although $c^2<1$ is preferred in our model, phenomenologically, $c^2\geq 1$ could also be obtained if we allow some modifications of the temperature or the entropy of the holographic gas. For example, one can take $S>\frac{a+3+m}{a+3}\times 8\pi^2R_h^2M_p^2$ or $T>\frac{a+3+m}{a+3}\times\frac{1}{2\pi R_h}$. However, these possibilities either break the de Sitter entropy bound, or require a higher temperature than the Gibbons-Hawking temperature. The underlying physics behind these modifications remains to be explained.

Alternatively, if the holographic gas obeys infinite statistics, then we show in Appendix C that $c^2$ approaches to 1 from below as the particle number increases. When the number of holographic particle is large,
\begin{equation}
  c^2\simeq 1~.
\end{equation}

The constant $\omega_0$ can be calculated by comparing the entropy in Eq. \eqref{ES} with the Gibbons-Hawking entropy,
\begin{align}
  &\omega_{0{\rm MB}}=\frac{6}{a+4}\frac{1}{\Gamma(a+3)}\left(\frac{3}{4}\right)^b 2^{a+4}\pi^{a-b+6}~,\\
  &\omega_{0{\rm BE}}=\frac{\omega_{0{\rm MB}}}{\zeta(a+4)}~, ~~~\omega_{0{\rm FD}}=\frac{\omega_{0{\rm MB}}}{\left(1-2^{-a-3}\right)\zeta(a+4)}~.
\end{align}

For example, when we take $a=1$ and $b=1$, we have $c^2=0.8$, $\omega_{0{\rm MB}}=24\pi^6/5$ and $S=5 \ln{\cal Z}$. When $a=-2$ and $b=0$, we have $c^2=0.5$, $\omega_{0{\rm MB}}=12\pi^4$, and $S=2\ln{\cal Z}$.

\subsection{From Energy Density to Evolution}
As we have shown, the energy density of holographic gas takes the form of holographic dark energy. So the cosmic evolution of holographic gas is the same as that of holographic dark energy. We only quote the result here \cite{Li:2004rb}. In a matter dominated universe, the relative energy density $\Omega_{\rm HG}=\rho/(3M_p^2H^2)$ satisfies the following differential equation,
\begin{equation}\label{omegahg}
\frac{\Omega_{\rm HG}'}{\Omega_{\rm HG}}=(1-\Omega_{\rm HG})(1+\frac{2\sqrt{\Omega_{\rm HG}}}{c})~,
\end{equation}
where prime denotes derivative with respect to $x\equiv \ln a$. The solution of Eq. \eqref{omegahg} can be written as
\begin{equation}\label{omegahg2}
\ln \Omega_{\rm HG}-\frac{c^2-2c}{c^2-4}\ln(1-\sqrt{\Omega_{\rm
HG}})-\frac{c^2+2c}{c^2-4}\ln(1+\sqrt{\Omega_{\rm
HG}})+\frac{8}{c^2-4}\ln(1+\frac{2}{c}\sqrt{\Omega_{\rm HG}})=\ln
a+x_0 ~.
\end{equation}
If we set $a_0=1$ at present time, $x_0$ is equal to the L.H.S. of
\eqref{omegahg2} with $\Omega_{\rm HG}$ replaced by $\Omega_{\rm
HG}^0$, which denotes the relative energy density of dark energy at
the present time.

The effective equation of state $w_{\rm eff}$ takes the form
\begin{equation}
  w_{\rm eff}=-\frac{1}{3}-\frac{2}{3}\frac{\sqrt{\Omega_{\rm HG}}}{c}~.
\end{equation}
When holographic gas does not dominate the total energy density,
$w_{\rm eff}\simeq -1/3$ and $\rho\propto a^{-2}$. When holographic
gas dominates the total energy density, holographic gas behaves like
phantom. As investigated in \cite{Li:2004rb, Chen:2006qy},
holographic gas also plays a role at the early stage of inflation.

\section{Conclusion}
To conclude, in this paper, we have investigated the statistical
nature of holographic gas.

We have calculated the grand partition function of holographic gas
in Maxwell-Boltzmann, Bose-Einstein, Fermi-Dirac and infinite
statistics. Thermodynamical quantities such as energy, entropy,
particle number, pressure and specific heat are obtained from the
grand partition function. We also verified the stability of
holographic gas.

We have applied holographic gas to dark energy. After a clarification on physical pressure and effective pressure, we applied the Gibbons-Hawking entropy and temperature to holographic gas. We find that the energy density reproduces holographic dark energy if holographic gas spreads inside the future event horizon. The parameter $c$ of holographic dark energy can also be calculated from holographic gas, which has been left as a free parameter before. The simplest holographic gas models predict $c^2<1$, which is in agreement with data fitting.

As a closing remark, we also would like to comment on the ``old
cosmological constant problem'' (why the original cosmological
constant is zero) in the holographic gas scenario. It seems that we
have assumed the vacuum energy to be zero before we write the
holographic gas energy density $\rho_{\rm HG}$ in the Friedmann
equation, so that the old cosmological constant problem remains
unsolved. However, one should keep in mind that the holographic gas
itself is already the quasi-particle description of gravity. So an
inclusion of the original cosmological constant should be a double
counting in our scenario.

As a comparison, the holographic dark energy scenario also does not suffer the old cosmological constant problem, but for a different reason. In the holographic dark energy scenario, the UV cutoff $\Lambda_{\rm UV}$ is holographicly related to the IR cutoff. So the quantum zero point energy $\rho_\Lambda\sim \Lambda_{\rm UV}^4$ becomes small. Again, the same result from different origins between holographic gas and holographic dark energy show evidence that there may be some nontrivial connections between the two.

\section*{Acknowledgments}
This work was supported by grants from NSFC, a grant from Chinese
Academy of Sciences, a grant from USTC, and a 973 project grant.

\section*{Appendix A: Holographic Gas with Modified Dispersion Relation}

In this Appendix we investigate holographic gas with modified
degeneracy and dispersion relation. For simplicity, we only consider
indistinguishable Maxwell-Boltzmann statistics. Bose-Einstein and
Fermi-Dirac statistics produce similar results.

In this Appendix, we assume that degeneracy and dispersion relation
of the gas is
\begin{equation}
  \omega=\omega_0 k^aV^bM_p^{3b-a}~,~~~\epsilon=M_p^{3n-m+1}k^mV^n~,
\end{equation}
We have
 \begin{equation}
\ln {\cal
Z}=Z_1=\frac{\omega_0}{2m\pi^2}\Gamma(\frac{a+3}{m})M_p^{3b-a-\frac{a+3}{m}(3n-m+1)}V^{1+b-\frac{(a+3)n}{m}}T^{\frac{a+3}{m}}
\end{equation}
We require $(a+3)/m>0$ to make the partition function finite. The
energy $E$ is automatically positive in this case.

We take $\mu=0$. The total energy and total entropy take the form
\begin{equation}\label{ESM}
  E=T^2\partial_T \ln {\cal Z}+\mu T \partial_\mu \ln {\cal Z}=\frac{a+3}{m}T\ln {\cal Z}~,~~~S=E/T+\ln {\cal Z}=(\frac{a+3}{m}+1)\ln {\cal Z}~.
\end{equation}

Inspired by holography, we take $T\propto V^{-1/3}$, and require
that $S$ is proportional to the area of the system. This requirement
is reduced to a relation between $a$, $b$, $m$ and $n$,
\begin{equation}
3b-\frac{(a+3)(3n+1)}{m}+1=0
\end{equation}

The energy density can be written as
\begin{equation}
  \rho=\frac{a+3}{a+3+m}\frac{ST}{V}~,
\end{equation}
and $(a+3)/m>0$ is required to make the integral finite.

\section*{Appendix B: Holographic Gas with Negative Thermodynamical Pressure}
As discussed in Sec. \ref{Stability}, although the holographic gas
produces negative effective pressure, the physical pressure is still
positive in the given examples. In this Appendix, we show that with
some assumptions on degeneracy and dispersion relation, gas with
negative pressure can be obtained. However, it is the effective
pressure, not the physical pressure, that directly relates with the
acceleration of expansion of the universe. So the results in this
appendix do not have direct implication for dark energy. For
simplicity, we only consider indistinguishable Maxwell-Boltzmann
statistics. Bose-Einstein and Fermi-Dirac statistics produce similar
results. We only consider the $\mu=0$ case for simplicity.

In this appendix, we assume that degeneracy and dispersion relation
of the gas is
\begin{equation}
  \omega=\omega_0 k^aV^bM^{3b-a}~,~~~\epsilon=M^{3n-m+1}k^mV^n~,
\end{equation}
where $M$ is a mass scale. We have
\begin{equation}
  \ln {\cal Z}=Z_1=\frac{\omega_0}{2m\pi^2}\Gamma(\frac{a+3}{m})M^{3b-a-\frac{a+3}{m}(3n-m+1)}V^{1+b-\frac{(a+3)n}{m}}T^{\frac{a+3}{m}}
\end{equation}
We require $(a+3)/m>0$ to make the partition function finite. The energy $E$ is automatically positive in this case.

With suitable values of $a$, $b$, $m$ and $n$, we can get negative physical pressure $p=T\partial_V \ln {\cal Z}$. To see this, we note that
\begin{equation}
  w=\frac{p}{\rho}=\frac{pV}{E}=\frac{(1+b)-(a+3)n/m}{(a+3)/m}~.
\end{equation}
So when $1+b-n(a+3)/m<0$, the pressure is negative. Particularly,
when $1+b=(a+3)(n-1)/m$ we have $p=-\rho$. In this case $3b-a<0$,
and $M$ should be small to make the degeneracy $\omega$ more than 1.

Gas with negative pressure usually suffers the problem of instability.
However, if the wavelength of the typical particles in the gas is
very long, then a large volume of negative pressure gas can be
realized. Although this negative physical pressure does not have
direct application to dark energy, it is interesting by its own
right.

\section*{Appendix C: Holographic Gas with Infinite Statistics }
It is suggested that dark energy may be explained by particles with
infinite statistics \cite{ng}. In this appendix we investigate this
idea in detail. We find that there will be a problem regarding
particle number if we assume Euler's relation in infinite
statistics. However, it is still suitable to use infinite statistics
to explain dark energy.

The infinite statistics assumes that, instead of $a_j a_k^\dagger \mp a_k^\dagger a_j =\delta_{jk}$ in the Bose-Einstein and Fermi-Dirac cases, the relations between the creation and annihilation operators takes the form
\begin{equation}
  a_j a_k^{\dagger}=\delta_{jk}~.
\end{equation}
No other commutation relation is available. For example,
$a_j^\dagger a_k^\dagger |0\rangle \neq a_k^\dagger a_j^\dagger
|0\rangle$, because $a_j (a_j^\dagger a_k^\dagger |0\rangle) \neq
a_j (a_k^\dagger a_j^\dagger |0\rangle)$. One can count states
to verify that the partition function of infinite statistics is just
that of the Maxwell-Boltzmann statistics with distinguishable particles.

We use $Z_1$ to denote the single particle canonical partition function. The grand partition can be written in terms of $Z_1$ as
\begin{equation}\label{ZIF}
  {\cal Z}=\sum_{N=0}^{\infty}\left(e^{\mu/T}Z_1\right)^N~.
\end{equation}

If $e^{\mu/T}Z_1>1$, $\cal Z$ diverges, and the particle number becomes infinite. So only $e^{\mu/T}Z_1<1$ makes sense in physics, and
\begin{equation}
  {\cal Z}=\frac{1}{1-e^{\mu/T}Z_1}~.
\end{equation}

The particle number, pressure and total entropy takes the form
\begin{equation}
  N={\cal Z}-1,~~~p=NT \partial_V \ln Z_1,~~~S=-{\mu}N/T+E/T+\ln {\cal Z}~.
\end{equation}
Assume Euler's Relation $TS=E+pV-{\mu}N$ we have $\ln
(N+1)=N(V\partial_V \ln Z_1)$, which means $N\sim 1$.

However, there is no Euler's Relation for Holographic gas, because
the entropy is proportional to square, not volume. Still, we assume
relativistic dispersion relation $\epsilon=k$ and the degeneracy
with the form $\omega =\omega_0 k^aV^bM_p^{3b-a}$. The same as
Maxwell-Boltzmann statistics, the single particle canonical
partition function is $Z_1
=(2\pi^2)^{-1}\omega_0\Gamma(a+3)M_p^{3b-a}V^{1+b}T^{a+3}$. We note
that when $\mu=0$ the total energy and total entropy takes the form

\begin{equation}
  E=(a+3)NT,~~~S=(a+3)N+\ln(N+1)
\end{equation}
The energy density can be written as
\begin{equation}
 \rho=\frac{E}{V}=(a+3)\frac{NT}{ V}=[1+\frac{\ln(N+1)}{N(a+3)}]^{-1}\frac{ST}{V}
\end{equation}
Again we use the Gibbons-Hawking entropy $S=8\pi^2R_h^2M_p^2$ and
temperature $T=1/(2\pi R_h)$, the energy density takes the form
\begin{equation}
\rho=3[1+\frac{\ln(N+1)}{N(a+3)}]^{-1} M_p^2R_h^{-2}
\end{equation}
Note that in our universe, the particle number $N$ is very large. So
to very good accuracy, we have $\rho=3M_p^2R_h^{-2}$. We conclude
that infinite statistics predicts $c\simeq 1$.

\end{document}